\documentclass[11pt]{article}

\usepackage[margin=1in]{geometry}
\usepackage{amsmath,amssymb,amsthm}
\usepackage{graphicx}
\usepackage{booktabs}
\usepackage{array}
\usepackage{algorithm}
\usepackage{algpseudocode}
\usepackage[hidelinks]{hyperref}
\usepackage[numbers,sort&compress]{natbib}

\newtheorem{theorem}{Theorem}
\newtheorem{lemma}{Lemma}
\newtheorem{corollary}{Corollary}
\newtheorem{proposition}{Proposition}
\theoremstyle{remark}

\newcommand{\E}{\mathbb{E}}
\renewcommand{\Pr}{\mathbb{P}}
\newcommand{\tr}{\operatorname{tr}}
\newcommand{\Cov}{\operatorname{Cov}}
\newcommand{\bary}{\bar y}
\newcommand{\bare}{\bar\varepsilon}

\title{Low-Rank Denoising for Predict-then-Optimize}

\author{Henry Aldridge-Krawciw\thanks{Regis High School, New York, NY 10028. haldridge29@regis.org}
\and Irene Aldridge\thanks{Cornell Tech, New York, NY 10044. irene.aldridge@gmail.com (corresponding author)}}
\date{}

\begin{document}
\maketitle

\begin{abstract}
\noindent A shortest-path decision is often made from edge costs that are only observed through noisy samples, such as the average of $m$ simulation replications. When the mean cost vectors of different instances lie near a $k$-dimensional subspace, projecting the averaged sample onto that subspace should remove noise. We make this precise. The mean squared error of any projection estimator splits exactly into a variance term and a bias term (Theorem~\ref{thm:mse}). The best projector is the positive eigenspace of $\Sigma_c - D/m$, where $\Sigma_c$ is the covariance of the true costs and $D$ is the noise covariance. For a rank-$k$ signal, the projection has the
error of a sample mean with a factor $d/(\mu k)$ more replications, where $d$ is the number of edges and $\mu$ is an incoherence constant. For a signal that is only approximately low-rank, every rank-$k$ projection has an error floor, and the plain average wins beyond an explicit replication count (Corollary~\ref{cor:floor}). A margin bound shows that the chance of choosing a suboptimal path falls at the same improved rate (Theorem~\ref{thm:regret}). A projector learned from $n$ noisy instances loses at most an $O(d/n)$ term (Theorem~\ref{thm:learned}). 
\end{abstract}


\section{Introduction}

Routing a vehicle, scheduling a sequence of tasks, and moving a packet through a network all reduce to finding a minimum-cost path \citep{vera2021,chen2004}. The graph is usually known. The edge costs, however, are not known. The edge costs are transit times, delays, or risks, and we see them only through noisy measurements or simulation runs \citep{nelson2021}. The standard remedy is to gather more samples and average them, and the error of that average falls only as $1/\sqrt m$ in $m$ samples. Costs on a network are not independent, though. Congestion, weather, and demand move many edges together, so the cost vectors of different instances are not spread evenly over all $d$ directions. This paper asks how much of the sampling effort that structure can replace and where it stops helping. We answer with exact error formulas, a rule for the best rank, and bounds on the chance of choosing a suboptimal path, and we test them on the standard shortest-path benchmark of \citet{elmachtoub2022}.

\subsection{Problem}

Let $G=(V,E)$ be a directed acyclic graph with $d=|E|$ edges and a set
$\mathcal Z\subseteq\{0,1\}^d$ of source-to-sink paths, each with at most $L$
edges. The true edge costs $c\in\mathbb{R}^d$ are unknown. For each instance, we see $m$ independent noisy replications $y^{(r)}=c+\varepsilon^{(r)}$ and form the average $\bary_m$. The decision is the shortest path under an estimate $\hat c$, and its quality is its \emph{regret} $c^\top\hat z-c^\top z^\ast$, where
$z^\ast$ is the optimal path. The default estimate is $\hat c=\bary_m$. Its error falls as $1/m$ in variance, so the number of replications $m$ is the budget.

Costs on a network are correlated. Congestion, weather and demand move many edges at once. If the mean cost vector $c$ varies across instances within a low-dimensional subspace, the components of $\bary_m$ outside that subspace are pure noise. This suggests the \emph{projection estimator}: learn the subspace by performing an eigenvalue decomposition (PCA) of past instances and projecting.
\emph{Predict-then-optimize} methods such as SPO+ \citep{elmachtoub2022} address a different problem in which costs are predicted from features and do not sample the current instance.

\subsection{Results}

We study the projection estimator $\hat c_\Pi=\bar c+\Pi(\bary_m-\bar c)$ for a rank-$k$ orthogonal projector $\Pi$, and its hybrid with a predictor. All results are for the setting above. We write $\Sigma_c=\Cov(c)$ and $D=\E[\Cov(\varepsilon\mid c)]$.

\begin{enumerate}
\item \textbf{Exact decomposition and optimal projector (Thm.~\ref{thm:mse}).}
$\E\|\hat c_\Pi-c\|^2=\tr(\Pi D)/m+\tr((I-\Pi)\Sigma_c)$. The projector that minimizes it over all ranks is the positive eigenspace of $\Sigma_c-D/m$. It is fully noise-aware and needs no hypothesis on the noise law. 
\item \textbf{Replication saving (Cor.~\ref{cor:saving}).} If $\operatorname{rank}\Sigma_c=k$, the error equals that of a sample mean with at least $m\,d/(\mu k)$ replications.
\item \textbf{Floor and crossover (Cor.~\ref{cor:floor}).} If $\operatorname{rank}\Sigma_c>k$, every rank-$k$ projection has an error of at least $\sum_{j>k}\lambda_j(\Sigma_c)$ as $m\to\infty$, and the sample mean is better once $m>\tr((I-\Pi)D)/\tr((I-\Pi)\Sigma_c)$.
\item \textbf{Regret with a margin (Thm.~\ref{thm:regret}).} With path gap $\Delta$, the probability of a suboptimal path is at most $2d\exp(-m\Delta^2 d/(8L^2b^2\mu k))$, an improvement of the exponent by $d/(\mu k)$ over the sample mean.
\item \textbf{Learned projector (Thm.~\ref{thm:learned}).} Estimating the subspace from $n$ noisy instances adds an error of order $d/n$ times a condition number.
\item \textbf{Hybrid (Prop.~\ref{prop:hybrid}).} Denoising the residual of a predictor $h$ improves on $h$ exactly when the noise kept by $\Pi$ is smaller than the residual signal it keeps.
\end{enumerate}

\subsection{Related work}

\paragraph{Learning to route.} Online shortest-path and combinatorial-bandit algorithms \citep{takimoto2003,awerbuch2004,gyorgy2007,cesabianchi2012}
minimize regret over a sequence of adversarial or stochastic cost vectors with bandit feedback. We instead have a one-shot decision with a batch of noisy samples from the \emph{current} instance and a history of past instances. The use of a learned structure to improve a decision is in the spirit of algorithms with predictions
\citep{mitzenmacher2022}.

\paragraph{Predict-then-optimize.} SPO+ \citep{elmachtoub2022} and related methods
\citep{bertsimas2020,ban2019,donti2017,wilder2019,kotary2021} learn costs from
features. Generalization and calibration bounds are known for SPO+
\citep{elbalghiti2023,liu2021}, and \citet{hu2022} show that plain
estimate-then-optimize can attain fast rates under a margin condition. Our margin argument (Theorem~\ref{thm:regret}) uses the same mechanism.

\paragraph{Low-rank estimation.} For a spiked covariance, sample eigenvectors are consistent above a phase-transition threshold \citep{baik2005,paul2007}, and optimal shrinkage and thresholding rules are known \citep{gavish2014,donoho2018}. However, \cite{gavish2014,donoho2018} analyses are for additive homoscedastic noise. Our decomposition is exact for noise that is multiplicative and heteroscedastic, because it uses only the second moment.
Perturbation of eigenspaces is controlled by the Davis--Kahan theorem
\citep{yu2015}, and covariance concentration by \citet{vershynin2018}. Randomized and Lanczos eigensolvers \citep{halko2011,lehoucq1998} give fast projections.

\paragraph{Simulation.} The average of $m$ replications and its confidence interval are the standard output of a stochastic simulation \citep{nelson2021}. Metamodels such as stochastic kriging \citep{ankenman2010} use structure across design points to reduce noise.
The projection estimator is a metamodel for a vector-valued output.

\section{Model}
\label{sec:model}

\paragraph{Instances and noise.} An instance has a random true cost vector $c\in\mathbb{R}^d$ with a mean $\bar c$ and covariance $\Sigma_c$. Given $c$, replications $\varepsilon^{(1)},\dots,\varepsilon^{(m)}$ are independent, with a mean of zero, independent coordinates, and $|\varepsilon^{(r)}_e|\le b$. Write $\bare=\frac1m\sum_r\varepsilon^{(r)}$ and $\bary_m=c+\bare$. Let $D=\E[\operatorname{diag}(\operatorname{Var}(\varepsilon_e\mid c))]$, a diagonal matrix. Then $\E[\bare\bare^\top]=D/m$. Noise may depend on $c$. In our experiments $\varepsilon_e=c_e(u_e-1)$ with $u_e\sim\mathrm{Uniform}[0.5,1.5]$, so that $D_{ee}=\E[c_e^2]/12$.

\paragraph{Estimators.} For a symmetric idempotent $\Pi$ of rank $k$ that is independent of the test noise,
\begin{equation}
\hat c_\Pi \;=\; \bar c+\Pi(\bary_m-\bar c).
\label{eq:proj}
\end{equation}
Taking $\Pi=I$ gives the sample mean. Let $\Delta(c)$ be the smallest cost difference between
$c^\top z^\ast$ and a path that is not optimal.

\paragraph{Experimental model.} Edge costs follow \citet{elmachtoub2022}:
$c_e(x)=[(Bx)_e/\sqrt p+3]^{\deg}$ with $x\sim N(0,I_p)$, $p=5$,
$B\in\{\pm1\}^{d\times p}$, on the $5\times5$ grid ($d=40$, $L=8$). The exponent $\deg\in\{1,2,4\}$
controls how far $c$ is from a rank-$p$ subspace.

\section{Results}
\label{sec:results}

\subsection{From error to regret}

\begin{lemma}
\label{lem:regret}
Let $\hat z\in\arg\min_{z\in\mathcal Z}\hat c^\top z$ and $z^\ast\in\arg\min_{z\in\mathcal Z}c^\top z$. Then
\[
c^\top\hat z-c^\top z^\ast\;\le\;(\hat c-c)^\top(z^\ast-\hat z)\;\le\;2L\|\hat c-c\|_\infty.
\]
If $\|\hat c-c\|_\infty<\Delta/(2L)$ the regret is zero.
\end{lemma}

\subsection{Mean squared error and the optimal projector}

\begin{lemma}
\label{lem:decomp}
Let $\Pi$ be an orthogonal projector independent of $\bare$, and $\bar c'\in\mathbb{R}^d$. For $\hat c=\bar c'+\Pi(\bary_m-\bar c')$,
\[
\hat c-c=\Pi\bare-(I-\Pi)(c-\bar c'),\qquad
\|\hat c-c\|^2=\|\Pi\bare\|^2+\|(I-\Pi)(c-\bar c')\|^2 .
\]
\end{lemma}

\begin{theorem}[Exact error and optimal projector]
\label{thm:mse}
For a fixed projector $\Pi$ and $\bar c'=\bar c$,
\begin{equation}
\E\|\hat c_\Pi-c\|^2\;=\;\frac{\tr(\Pi D)}{m}+\tr\big((I-\Pi)\Sigma_c\big)\;=\;\tr\Sigma_c-\tr\!\Big(\Pi\big(\Sigma_c-\tfrac Dm\big)\Big).
\label{eq:mse}
\end{equation}
Among all projectors, this is minimized by the projector onto the positive eigenspace of $\Sigma_c-D/m$. The minimum is
$\tr\Sigma_c-\sum_j(\lambda_j(\Sigma_c-D/m))_+$, which is at most $\min(\tr D/m,\ \tr\Sigma_c)$. Among rank-$k$ projectors, the minimizer is the top-$k$ eigenspace of $\Sigma_c-D/m$. In particular, the optimal rank
\[
k^\ast(m)=\#\{j:\lambda_j(\Sigma_c-D/m)>0\}
\]
is non-decreasing in $m$. If $D=\sigma^2I$, the optimal projector is the span of the eigenvectors of the observed covariance
$\Sigma_c+\sigma^2I/m$ whose eigenvalue exceeds $2\sigma^2/m$.
\end{theorem}

The optimal rank is determined by the noise covariance as a hard threshold on the observed spectrum equal to twice the noise variance.

\begin{corollary}[Replication saving]
\label{cor:saving}
Suppose $\operatorname{rank}\Sigma_c=k$ and $\Pi$ project onto the range of $\Sigma_c$. Let $\mu=\frac dk\max_e\Pi_{ee}\in[1,d/k]$. Then
\[
\E\|\hat c_\Pi-c\|^2=\frac{\tr(\Pi D)}m\le\frac{\mu k}{d}\cdot\frac{\tr D}{m},
\]
with equality $\frac kd\cdot\frac{\tr D}{m}$ when $D=\sigma^2I$. So $\hat c_\Pi$ has the error of a sample mean computed from at least $m\,d/(\mu k)$ replications.
\end{corollary}

\begin{corollary}[Floor and crossover]
\label{cor:floor}
For every rank-$k$ projector, as $m\to\infty$,
$\E\|\hat c_\Pi-c\|^2\ge\tr((I-\Pi)\Sigma_c)\ge\sum_{j>k}\lambda_j(\Sigma_c)$,
with equality in the second step for the top-$k$ eigenspace of $\Sigma_c$. The sample mean has error $\tr D/m\to0$. The projection beats the sample mean exactly when
\[
m\;<\;m^\ast(\Pi)\;:=\;\frac{\tr((I-\Pi)D)}{\tr((I-\Pi)\Sigma_c)}
\]
($m^\ast=\infty$ if the denominator is zero). For $D=\sigma^2I$ and the top-$k$ eigenspace, $m^\ast=(d-k)\sigma^2/\sum_{j>k}\lambda_j(\Sigma_c)$.
\end{corollary}

An error floor is therefore intrinsic to a fixed rank and is not an artifact of a rank rule. A rank that grows with $m$, as $k^\ast(m)$ does, removes it.

\subsection{Regret with a margin}

\begin{theorem}[Suboptimality probability]
\label{thm:regret}
Let $\Pi$ be a rank-$k$ projector independent of the test noise, with $\rho=\max_e\Pi_{ee}$. Assume $(I-\Pi)(c-\bar c)=0$. Then for every $t>0$,
\[
\Pr\big(\|\hat c_\Pi-c\|_\infty\ge t\mid c\big)\le2d\exp\!\Big(-\frac{m t^2}{2b^2\rho}\Big),
\qquad
\Pr(\text{regret}>0\mid c)\le2d\exp\!\Big(-\frac{m\Delta^2}{8L^2b^2\rho}\Big).
\]
With $\rho\le\mu k/d$ the exponent exceeds that of the sample mean ($\rho=1$) by a factor $d/(\mu k)$. If instead $\beta=\|(I-\Pi)(c-\bar c)\|_2\le\Delta/(4L)$, then $\Pr(\text{regret}>0\mid c)\le2d\exp(-m\Delta^2/(32L^2b^2\rho))$.
\end{theorem}

The bound is a worst-case guarantee. It says that regret is zero with high probability once the error is below the margin, and that the improvement is the same factor $d/(\mu k)$ as in the mean squared error.

\subsection{Learning the subspace}

\begin{theorem}[Sample complexity]
\label{thm:learned}
Let $\operatorname{rank}\Sigma_c=k$ with eigenvalues $\lambda_1\ge\dots\ge\lambda_k>0$ and noise covariance $\sigma^2I/m=:s^2I$. Let the centered training vectors be sub-Gaussian with $\|\cdot\|_{\psi_2}\le K\|\Sigma_y\|^{1/2}$, where $\Sigma_y=\Sigma_c+s^2I$. Let $\hat\Pi$ project onto the top-$k$ eigenvectors of the sample covariance of $n\ge d$ training instances and $\hat c$ be the training mean. With probability at least $1-2e^{-d}$ over the training set, for a fresh instance,
\[
\E\big[\|\hat c_{\hat\Pi}-c\|^2\mid\text{train}\big]\;\le\;k\,s^2+C\,K^4\,\frac{(\lambda_1+s^2)^2\,\tr\Sigma_c}{\lambda_k^2}\cdot\frac dn+2\|\hat c-\bar c\|^2,
\]
with $\E\|\hat c-\bar c\|^2=\tr\Sigma_y/n$, and $C$ an absolute constant.
\end{theorem}

The first term is the oracle-subspace error $k s^2$ from Corollary~\ref{cor:saving}. The learned subspace costs a term of order $d/n$, so $n$ of order $d$ times a condition number suffices to be within a constant factor of the oracle. The constant $C$ is not tracked, and the bound is conservative: in our experiments, the loss is small already at $n/d=5$ (Section~\ref{sec:exp}). The phase-transition threshold for detecting a spike is $\lambda_k>s^2\sqrt{d/n}$ \citep{baik2005,paul2007}. It is far below the signal eigenvalues in our experiments at every $n/d\ge1.25$, so detectability is not what limits the rank estimate there (Section~\ref{sec:exp}).

\subsection{Hybrid with a predictor}

Let $h(x)$ be any predictor (for example, SPO+) and $r=c-h(x)$ its residual, with mean $\bar r$ and covariance $\Sigma_r$. The hybrid is
$\hat c^{\rm hyb}=h(x)+\bar r+\Pi(\bary_m-h(x)-\bar r)$.

\begin{proposition}[Hybrid]
\label{prop:hybrid}
For a fixed projector $\Pi$,
\[
\E\|\hat c^{\rm hyb}-c\|^2=\frac{\tr(\Pi D)}m+\tr((I-\Pi)\Sigma_r),
\]
while the predictor alone, $h(x)+\bar r$, has error $\tr\Sigma_r$. The hybrid is better iff $\tr(\Pi D)/m<\tr(\Pi\Sigma_r)$: the noise that $\Pi$ lets through is smaller than the residual signal it keeps. With $\Pi$ the positive eigenspace of $\Sigma_r-D/m$ the hybrid is never worse than either the predictor or the sample mean.
\end{proposition}

\section{Computation}
\label{sec:comp}

The oracle is a dynamic program over a topological order of the DAG and takes $O(d)$ time per cost vector. It accepts negative costs, which occur in the SPO+ update ($2\hat c-c$) and in denoised costs. Dijkstra's algorithm does not, so we use the dynamic program throughout. Fitting the projector requires the covariance of $n$ vectors in $O(nd^2)$ time and a full eigendecomposition in $O(d^3)$. A truncated decomposition of the $n\times d$ data matrix, by Lanczos or randomized methods \citep{halko2011,lehoucq1998}, takes $O(ndk)$. Applying the projector costs $O(dk)$ per instance. In our timing test, the truncated solver was $253\times$ faster than a full eigendecomposition at $d=3120$ (Table~\ref{tab:timing} in the appendix), and the subspaces agreed to about $10^{-8}$. SPO+ training makes $O(nE)$ oracle calls for $E$ epochs; that is, $O(nEd)$ time.

\section{Experiments}
\label{sec:exp}

All experiments use the model of Section~\ref{sec:model}. Each experiment replication draws a fresh $B$, a training set, and a test set, and all methods see the same data. Tests are paired across experiment replications. SPO+ is the method of \citet{elmachtoub2022}: a linear predictor $\hat c(x)=Wx$ fitted by minimizing the SPO+ loss with stochastic subgradient descent, $W\leftarrow W-\eta_t\,2(z^\ast-z_{\rm spo})x^\top$ plus an $\ell_2$ penalty of $10^{-4}$, where $z_{\rm spo}$ is the shortest path under $2Wx-c$. The step size is $\eta_t=\eta_0/\sqrt t$ in epoch $t$. The starting value $\eta_0\in\{0.1,0.02,0.005,0.001\}$ is chosen on a $20\%$ hold-out by the mean observed cost of the path selected, and the model is then refitted on all data. This uses no true costs. Table~\ref{tab:main} also reports SPO+ with the fixed step size $0.02$ as a sensitivity check.

\subsection{The error formulas}

Table~\ref{tab:thm} compares the measured error ratio (projection over sample mean) with the population formula of Theorem~\ref{thm:mse}, $200$ experiment replications, $n=200$. 

\begin{table}[ht]
\centering
\caption{Mean squared error of the GD-rank projection divided by that of the sample mean ($m$ replications per instance). Theory uses the population $\Sigma_c$ and $D$, with $k=\hat k$ or the optimal rank $k^\ast(m)$ of Theorem~\ref{thm:mse}.}
\label{tab:thm}
\footnotesize
\begin{tabular}{@{}llrrrrrr@{}}
\toprule
deg & $m$ & Measured & Theory, $k=\hat k$ & Theory, optimal & $\hat k$ (GD) & $k^\ast$ \\
\midrule
1 & 1 & 0.16 & 0.13 & 0.13 & 5.0 & 5.0 \\
 & 8 & 0.16 & 0.13 & 0.13 & 5.0 & 5.0 \\
 & 64 & 0.17 & 0.14 & 0.14 & 5.0 & 5.0 \\
\midrule
2 & 1 & 0.31 & 0.28 & 0.27 & 5.1 & 6.2 \\
 & 8 & 0.62 & 0.52 & 0.39 & 11.3 & 15.0 \\
 & 64 & 0.49 & 0.40 & 0.40 & 15.7 & 16.0 \\
\midrule
4 & 1 & 0.77 & 0.60 & 0.45 & 12.1 & 15.8 \\
 & 8 & 1.10 & 0.81 & 0.55 & 15.8 & 21.3 \\
 & 64 & 4.33 & 2.84 & 0.57 & 16.6 & 22.6 \\
\bottomrule
\end{tabular}

\end{table}

For linear costs ($\deg=1$, $\operatorname{rank}\Sigma_c=5$), the formula gives $k/d=0.125$ and we measure $0.156$. The gap is the cost of learning the subspace from $200$ instances. Corollary~\ref{cor:saving} is then a saving of $8\times$ in theory and $6.4\times$ in measurement; the incoherence constant, averaged over draws of $B$, is $\mu=1.44$, which guarantees at least $5.6\times$. For $\deg=2$ and $\deg=4$, the formula with $k=\hat k$ tracks the measured ratio and its rise with $m$, and predicts the crossover of Corollary~\ref{cor:floor}: $m^\ast\approx5.8$ for $k=5$ at $\deg=2$, and between $2.4$ and $12.6$ for $k=12$–$16$ at $\deg=4$. The measured ratio at $\deg=4$ passes $1$ between $m=4$ and $m=8$, inside that range.

\subsection{The optimal-rank rule and the floor}

The Gavish--Donoho (GD) rule \citep{gavish2014} assumes homoscedastic noise, which this design violates, and its rank grows more slowly with $m$ than $k^\ast(m)$ does. We compare it with the rule of Theorem~\ref{thm:mse}, implemented with a plug-in noise covariance (the noise family is known, so $D_{ee}/m=\E[y_e^2]/(12m+1)$ is estimated from the training mean squares). Table~\ref{tab:t3} gives mean regret.

\begin{table}[ht]
\centering
\caption{Mean regret (\%) and selected rank, $n=200$, $200$ experiment replications. ``Theorem rank'' uses the positive eigenspace of $\hat\Sigma_y-2\hat D/m$.}
\label{tab:t3}
\footnotesize

\begin{tabular}{@{}llrrrrr@{}}
\toprule
deg & $m$ & Sample mean & GD rank & Theorem~\ref{thm:mse} rank & $\hat k$ (GD) & $\hat k$ (Thm.) \\
\midrule
1 & 1 & 6.64 & 1.50 & 1.50 & 5.0 & 5.0 \\
 & 8 & 1.11 & 0.19 & 0.19 & 5.0 & 5.1 \\
 & 64 & 0.15 & 0.03 & 0.03 & 5.0 & 5.3 \\
\midrule
2 & 1 & 5.80 & 1.89 & 2.22 & 5.1 & 8.4 \\
 & 8 & 0.75 & 0.56 & 0.44 & 11.3 & 14.8 \\
 & 64 & 0.09 & 0.05 & 0.05 & 15.7 & 16.0 \\
\midrule
4 & 1 & 4.58 & 6.33 & 5.23 & 12.1 & 15.8 \\
 & 8 & 0.53 & 1.50 & 0.85 & 15.8 & 20.5 \\
 & 64 & 0.07 & 0.90 & 0.08 & 16.6 & 22.9 \\
\bottomrule
\end{tabular}

\end{table}

At $\deg=4$ the GD rank stops growing near $16$ and regret reaches a floor of $0.90\%$ at $m=64$, while the sample mean reaches $0.07\%$. The theorem rank keeps growing and reaches $0.08\%$, so the floor disappears. At $\deg=2$, the theorem rank is better from $m=4$ on (for example, $0.83\%$ against $0.95\%$ at $m=4$, a paired difference of $0.12\pm0.03$). At $m=1$, however, regret is worse than with GD at $\deg=2$ ($2.22\%$ against $1.89\%$) and worse than the sample mean at $\deg=4$ ($5.23\%$ against $4.58\%$), although the mean squared error is lower. Mean squared error is not regret: the projection can distort the gaps between paths. Theorem~\ref{thm:regret} is a statement about the error in the maximum norm, which the optimal-MSE rule does not minimize.

\begin{figure}[ht]
\centering
\includegraphics[width=\textwidth]{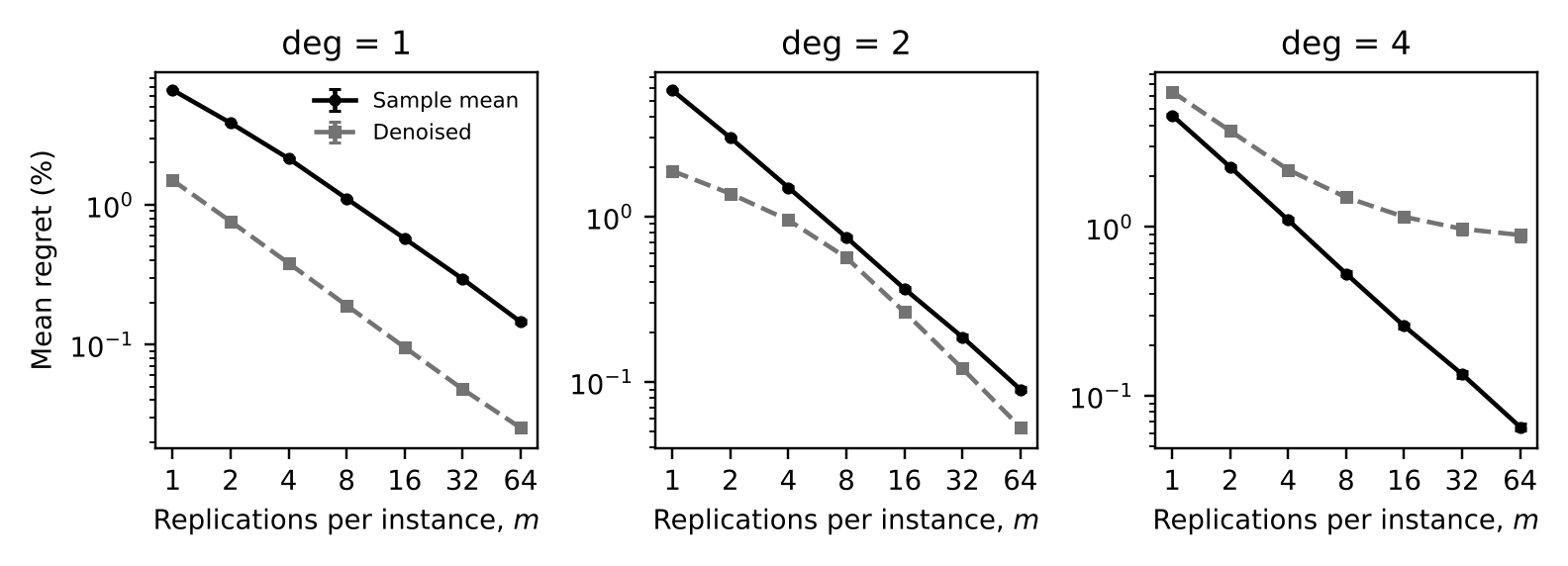}
\caption{Mean regret against $m$ for the sample mean and the GD-rank projection, with $95\%$ confidence intervals. At $\deg=1$ one projected sample matches about six replications. At $\deg=4$ the projection has a floor.}
\label{fig:simbudget}
\end{figure}

\subsection{Comparison with Existing Methods}
\label{sec:spo}

Table~\ref{tab:main} compares methods at $m=1$ with SPO+, PCA with different $k$ selections, Ledoit-Wolf filter, and hybrid SPO+/PCA models, where the data is pre-processed using PCA before being fed to SPO+. All experiments were conducted on $100$ replications. More extensive results are presented in the Appendix.

\begin{table}[ht]
\centering
\caption{Mean regret (\%), $m=1$, $n=200$, $d=40$, $100$ experiment replications, means $\pm$ $95\%$ half-widths. Best per column in bold.}
\label{tab:main}
\small
\begin{tabular}{@{}lccc@{}}
\toprule
Method & $\deg=1$ & $\deg=2$ & $\deg=4$ \\
\midrule

Noisy (naive Dijkstra) & 6.71 $\pm$ 0.10 & 5.87 $\pm$ 0.11 & 4.57 $\pm$ 0.11 \\
SPO+, fixed step size (sensitivity) & 1.58 $\pm$ 0.08 & 1.38 $\pm$ 0.07 & 7.50 $\pm$ 0.45 \\
SPO+ & 0.82 $\pm$ 0.08 & 1.38 $\pm$ 0.09 & 8.04 $\pm$ 0.49 \\
PCA, $k{=}5$ & 1.49 $\pm$ 0.03 & 1.87 $\pm$ 0.06 & 12.38 $\pm$ 0.59 \\
PCA, GD $\hat k$ & 1.49 $\pm$ 0.03 & 1.87 $\pm$ 0.06 & 6.17 $\pm$ 0.29 \\
Ledoit-Wolf filter & 1.75 $\pm$ 0.03 & 1.77 $\pm$ 0.05 & \textbf{2.91 $\pm$ 0.10} \\
Hybrid, $k{=}5$ & 0.57 $\pm$ 0.03 & \textbf{1.24 $\pm$ 0.07} & 10.61 $\pm$ 0.58 \\
Hybrid, GD $\hat k$ & \textbf{0.57 $\pm$ 0.03} & 1.24 $\pm$ 0.07 & 6.47 $\pm$ 0.31\\

\bottomrule
\end{tabular}
\end{table}

The hybrid PCA/SPO+ is best at $\deg=1$ ($0.57\%$, $p<0.001$ against both) and at $\deg=2$ ($1.24\%$ against $1.38\%$). Proposition~\ref{prop:hybrid} states when this can happen: the projection must retain more residual signal than noise. We did not measure the residual spectrum, so we do not claim that this is the reason. The advantage shrinks with $n$ (to $0.009\%$ against $0.029\%$ at $n=10^5$) and reverses at $\deg=2,4$ for large $n$ (appendix). 

\section{Discussion and Limitations}

The theory explains when and why denoising helps: the saving factor $d/(\mu k)$ for a low-rank signal, the floor for a signal that is not, the crossover count $m^\ast$, and an optimal rank rule. Its limits are as follows. The regret theorem is a worst-case bound in the maximum norm, with the constant $2L$ coming from path length. The learned-subspace theorem assumes homoscedastic noise and an exactly low-rank signal, and its constants are not tracked. Corollary~\ref{cor:saving} needs $\Sigma_c$ exactly rank $k$, while in the experiments $\deg=1$ is exactly rank five up to a clipping effect, and the others are not. Theorem~\ref{thm:mse} optimizes squared error, not regret, and Section~\ref{sec:exp} shows the two can disagree at $m=1$. All experiments use one synthetic family with multiplicative uniform noise on small networks and no real data. 

\section{Conclusions}

We studied the projection of an averaged cost sample onto a learned low-rank subspace as a way to make better shortest-path decisions from noisy edge costs. The main findings are as follows.

\begin{enumerate}
\item The error of any projection estimator splits exactly into a noise term and a bias term (Theorem~\ref{thm:mse}). The best projector is the positive eigenspace of $\Sigma_c-D/m$, so the best rank is set by the noise covariance and grows with the number of replications.
\item For an exactly rank-$k$ signal, the projection is worth a factor of at least $d/(\mu k)$ in replications. We measured $6.4\times$ on the $5\times5$ grid against $8\times$ in theory (Corollary~\ref{cor:saving}).
\item For a signal that is not exactly low-rank, every fixed rank has an error floor, and the plain average wins beyond an explicit replication count (Corollary~\ref{cor:floor}). A rank that grows with $m$ removes the floor (regret $0.08\%$ against $0.90\%$ at $m=64$ for the strongest nonlinearity).
\item The probability of a suboptimal path falls at the improved rate under a margin condition (Theorem~\ref{thm:regret}), and a projector learned from $n$ instances costs an additional $O(d/n)$ term (Theorem~\ref{thm:learned}).
\item Denoising the residual of a predictor helps when the noise retained by the projector is smaller than the residual signal it retains (Proposition~\ref{prop:hybrid}). Against SPO+, the hybrid has regret $0.57\%$ against $0.82\%$ at linear costs.
\end{enumerate}

\bigskip
\noindent\textbf{Code and data.} All experiments are simulations. 

\bibliographystyle{plainnat}
\bibliography{denoising_references}

@article{ban2019,
  author  = {Ban, Gah-Yi and Rudin, Cynthia},
  title   = {The Big Data Newsvendor: Practical Insights from Machine Learning},
  journal = {Operations Research},
  year    = {2019},
  volume  = {67},
  number  = {1},
  pages   = {90--108}
}

@article{bertsimas2020,
  author  = {Bertsimas, Dimitris and Kallus, Nathan},
  title   = {From Predictive to Prescriptive Analytics},
  journal = {Management Science},
  year    = {2020},
  volume  = {66},
  number  = {3},
  pages   = {1025--1044}
}

@inproceedings{chen2004,
  author    = {Chen, Shigang and Song, Meongchul and Sahni, Sartaj},
  title     = {Two Techniques for Fast Computation of Constrained Shortest Paths},
  booktitle = {Proceedings of the IEEE Global Telecommunications Conference (GLOBECOM '04)},
  year      = {2004},
  volume    = {3},
  pages     = {1348--1352},
  address   = {Dallas, Texas, USA},
  month     = {November 29--December 3}
}

@inproceedings{donti2017,
  author    = {Donti, Priya and Amos, Brandon and Kolter, J. Zico},
  title     = {Task-Based End-to-End Model Learning in Stochastic Optimization},
  booktitle = {Proceedings of the 31st Conference on Neural Information Processing Systems (NeurIPS)},
  year      = {2017},
  volume    = {30},
  address   = {Long Beach, California, USA},
  month     = {December 4--9}
}

@article{elmachtoub2022,
  author  = {Elmachtoub, Adam N. and Grigas, Paul},
  title   = {Smart ``{P}redict, then {O}ptimize''},
  journal = {Management Science},
  year    = {2022},
  volume  = {68},
  number  = {1},
  pages   = {9--26}
}

@article{gavish2014,
  author  = {Gavish, Matan and Donoho, David L.},
  title   = {The Optimal Hard Threshold for Singular Values is $4/\sqrt{3}$},
  journal = {IEEE Transactions on Information Theory},
  year    = {2014},
  volume  = {60},
  number  = {8},
  pages   = {5040--5053}
}

@inproceedings{kotary2021,
  author    = {Kotary, James and Fioretto, Ferdinando and Van Hentenryck, Pascal and Wilder, Bryan},
  title     = {End-to-End Constrained Optimization Learning: A Survey},
  booktitle = {Proceedings of the 30th International Joint Conference on Artificial Intelligence (IJCAI-21)},
  year      = {2021},
  pages     = {4475--4482},
  address   = {Montreal, Canada (virtual)},
  month     = {August 19--26}
}

@article{vera2021,
  author  = {Vera, Alberto and Banerjee, Siddhartha and Samaranayake, Samitha},
  title   = {Computing Constrained Shortest-Paths at Scale},
  journal = {Operations Research},
  year    = {2021},
  volume  = {70}
}

@inproceedings{wilder2019,
  author    = {Wilder, Bryan and Dilkina, Bistra and Tambe, Milind},
  title     = {Melding the Data-Decisions Pipeline: Decision-Focused Learning for Combinatorial Optimization},
  booktitle = {Proceedings of the 33rd AAAI Conference on Artificial Intelligence},
  year      = {2019},
  pages     = {1658--1665},
  address   = {Honolulu, Hawaii, USA},
  month     = {January 27--February 1}
}

@book{lehoucq1998,
  author    = {Lehoucq, Richard B. and Sorensen, Danny C. and Yang, Chao},
  title     = {{ARPACK} Users' Guide: Solution of Large-Scale Eigenvalue Problems with Implicitly Restarted {Arnoldi} Methods},
  publisher = {SIAM},
  year      = {1998}
}

@book{nelson2021,
  author    = {Nelson, Barry L. and Pei, Linda},
  title     = {Foundations and Methods of Stochastic Simulation: A First Course},
  publisher = {Springer},
  address   = {Cham},
  year      = {2021}
}

@article{ankenman2010,
  author  = {Ankenman, Bruce and Nelson, Barry L. and Staum, Jeremy},
  title   = {Stochastic Kriging for Simulation Metamodeling},
  journal = {Operations Research},
  volume  = {58},
  number  = {2},
  pages   = {371--382},
  year    = {2010}
}

@article{takimoto2003,
  author  = {Takimoto, Eiji and Warmuth, Manfred K.},
  title   = {Path Kernels and Multiplicative Updates},
  journal = {Journal of Machine Learning Research},
  volume  = {4},
  pages   = {773--818},
  year    = {2003}
}

@inproceedings{awerbuch2004,
  author    = {Awerbuch, Baruch and Kleinberg, Robert D.},
  title     = {Adaptive Routing with End-to-End Feedback: Distributed Learning and Geometric Approaches},
  booktitle = {Proceedings of the 36th ACM Symposium on Theory of Computing (STOC)},
  year      = {2004}
}

@article{gyorgy2007,
  author  = {Gy{\"o}rgy, Andr{\'a}s and Linder, Tam{\'a}s and Lugosi, G{\'a}bor and Ottucs{\'a}k, Gy{\"o}rgy},
  title   = {The On-Line Shortest Path Problem under Partial Monitoring},
  journal = {Journal of Machine Learning Research},
  volume  = {8},
  pages   = {2369--2403},
  year    = {2007}
}

@article{cesabianchi2012,
  author  = {Cesa-Bianchi, Nicol{\`o} and Lugosi, G{\'a}bor},
  title   = {Combinatorial Bandits},
  journal = {Journal of Computer and System Sciences},
  volume  = {78},
  number  = {5},
  pages   = {1404--1422},
  year    = {2012}
}

@article{elbalghiti2023,
  author  = {El Balghiti, Othman and Elmachtoub, Adam N. and Grigas, Paul and Tewari, Ambuj},
  title   = {Generalization Bounds in the Predict-then-Optimize Framework},
  journal = {Mathematics of Operations Research},
  volume  = {48},
  number  = {4},
  year    = {2023}
}

@inproceedings{liu2021,
  author    = {Liu, Heyuan and Grigas, Paul},
  title     = {Risk Bounds and Calibration for a Smart Predict-then-Optimize Method},
  booktitle = {Advances in Neural Information Processing Systems (NeurIPS)},
  volume    = {34},
  year      = {2021}
}

@article{hu2022,
  author  = {Hu, Yichun and Kallus, Nathan and Mao, Xiaojie},
  title   = {Fast Rates for Contextual Linear Optimization},
  journal = {Management Science},
  volume  = {68},
  number  = {6},
  pages   = {4236--4245},
  year    = {2022}
}

@article{baik2005,
  author  = {Baik, Jinho and Ben Arous, G{\'e}rard and P{\'e}ch{\'e}, Sandrine},
  title   = {Phase Transition of the Largest Eigenvalue for Nonnull Complex Sample Covariance Matrices},
  journal = {Annals of Probability},
  volume  = {33},
  number  = {5},
  pages   = {1643--1697},
  year    = {2005}
}

@article{paul2007,
  author  = {Paul, Debashis},
  title   = {Asymptotics of Sample Eigenstructure for a Large Dimensional Spiked Covariance Model},
  journal = {Statistica Sinica},
  volume  = {17},
  number  = {4},
  pages   = {1617--1642},
  year    = {2007}
}

@book{vershynin2018,
  author    = {Vershynin, Roman},
  title     = {High-Dimensional Probability: An Introduction with Applications in Data Science},
  publisher = {Cambridge University Press},
  year      = {2018}
}

@article{yu2015,
  author  = {Yu, Yi and Wang, Tengyao and Samworth, Richard J.},
  title   = {A Useful Variant of the {Davis--Kahan} Theorem for Statisticians},
  journal = {Biometrika},
  volume  = {102},
  number  = {2},
  pages   = {315--323},
  year    = {2015}
}

@article{donoho2018,
  author  = {Donoho, David L. and Gavish, Matan and Johnstone, Iain M.},
  title   = {Optimal Shrinkage of Eigenvalues in the Spiked Covariance Model},
  journal = {Annals of Statistics},
  volume  = {46},
  number  = {4},
  pages   = {1742--1778},
  year    = {2018}
}

@article{halko2011,
  author  = {Halko, Nathan and Martinsson, Per-Gunnar and Tropp, Joel A.},
  title   = {Finding Structure with Randomness: Probabilistic Algorithms for Constructing Approximate Matrix Decompositions},
  journal = {SIAM Review},
  volume  = {53},
  number  = {2},
  pages   = {217--288},
  year    = {2011}
}

@article{mitzenmacher2022,
  author  = {Mitzenmacher, Michael and Vassilvitskii, Sergei},
  title   = {Algorithms with Predictions},
  journal = {Communications of the ACM},
  volume  = {65},
  number  = {7},
  pages   = {33--35},
  year    = {2022}
}

@article{hoeffding1963,
  author  = {Hoeffding, Wassily},
  title   = {Probability Inequalities for Sums of Bounded Random Variables},
  journal = {Journal of the American Statistical Association},
  volume  = {58},
  number  = {301},
  pages   = {13--30},
  year    = {1963}
}

\appendix

\section{Proofs}

\begin{proof}[Proof of Lemma~\ref{lem:regret}]
Since $\hat c^\top\hat z\le\hat c^\top z^\ast$, we have $c^\top\hat z-c^\top z^\ast\le(\hat c-c)^\top(z^\ast-\hat z)$ by adding $(c-\hat c)^\top\hat z-(c-\hat c)^\top z^\ast$ on both sides. Each path has at most $L$ edges, so $\|z^\ast-\hat z\|_1\le2L$ and the second inequality follows. Regret is $0$ or at least $\Delta$. If $2L\|\hat c-c\|_\infty<\Delta$ the bound is below $\Delta$, so the regret is $0$.
\end{proof}

\begin{proof}[Proof of Lemma~\ref{lem:decomp}]
$\hat c-c=\bar c'+\Pi(c+\bare-\bar c')-c=\Pi\bare-(I-\Pi)(c-\bar c')$. The two terms lie in $\operatorname{range}\Pi$ and $\operatorname{range}(I-\Pi)$, which are orthogonal, so squared norms add.
\end{proof}

\begin{proof}[Proof of Theorem~\ref{thm:mse}]
By Lemma~\ref{lem:decomp}, $\E\|\hat c_\Pi-c\|^2=\E\tr(\Pi\bare\bare^\top)+\E\|(I-\Pi)(c-\bar c)\|^2$. Conditioning on $c$ and using independence and zero mean of the replications, $\E[\bare\bare^\top\mid c]=\operatorname{diag}(\operatorname{Var}(\varepsilon_e\mid c))/m$, whose expectation is $D/m$. The second term is $\tr((I-\Pi)\Sigma_c)$. Their sum is \eqref{eq:mse}. For a symmetric $A$ and orthogonal projector $\Pi$, $\tr(\Pi A)=\sum_jw_j\lambda_j(A)$ with $w_j=u_j^\top\Pi u_j\in[0,1]$ and $\sum_jw_j=\operatorname{rank}\Pi$. Take $A=D/m-\Sigma_c$. Over all projectors the minimum puts $w_j=1$ exactly on negative eigenvalues of $A$, which are the positive eigenvalues of $\Sigma_c-D/m$. Over rank-$k$ projectors the minimum is the sum of the $k$ smallest eigenvalues of $A$ (Ky Fan). The value follows from \eqref{eq:mse}. It is at most $\tr D/m$ ($\Pi=I$) and at most $\tr\Sigma_c$ ($\Pi=0$). Monotonicity in $m$ holds because $\Sigma_c-D/m$ is increasing in $m$ in the Loewner order. For $D=\sigma^2I$, $\Sigma_c-\sigma^2I/m$ and $\Sigma_y-2\sigma^2I/m$ coincide, where $\Sigma_y=\Sigma_c+\sigma^2I/m$.
\end{proof}

\begin{proof}[Proof of Corollary~\ref{cor:saving}]
Almost surely $c-\bar c\in\operatorname{range}\Sigma_c=\operatorname{range}\Pi$, so the bias term in \eqref{eq:mse} vanishes. As $D$ is diagonal and nonnegative, $\tr(\Pi D)=\sum_e\Pi_{ee}D_{ee}\le\max_e\Pi_{ee}\tr D$, and $\max_e\Pi_{ee}=\mu k/d$. If $D=\sigma^2I$ then $\tr(\Pi D)=k\sigma^2$. The sample mean with $m'$ replications has error $\tr D/m'$, which equals the bound when $m'=md/(\mu k)$.
\end{proof}

\begin{proof}[Proof of Corollary~\ref{cor:floor}]
The first inequality drops the nonnegative variance term. The second is Ky Fan: $\min_{\operatorname{rank}\Pi=k}\tr((I-\Pi)\Sigma_c)=\sum_{j>k}\lambda_j(\Sigma_c)$. The projection beats the mean iff $\tr(\Pi D)/m+T<\tr D/m$ with $T=\tr((I-\Pi)\Sigma_c)$, that is iff $T<\tr((I-\Pi)D)/m$.
\end{proof}

\begin{proof}[Proof of Theorem~\ref{thm:regret}]
\sloppy
By Lemma~\ref{lem:decomp} and the assumption, $\hat c_\Pi-c=\Pi\bare$. Coordinate $e$ is
\[
(\Pi\bare)_e=\frac1m\sum_{r}\sum_j\Pi_{ej}\varepsilon^{(r)}_j,
\]
a sum of independent terms (given $c$) with mean zero, the term for $(r,j)$ lying in $[-|\Pi_{ej}|b/m,\,|\Pi_{ej}|b/m]$. Hoeffding's inequality \citep{hoeffding1963} gives
\begin{align*}
\Pr(|(\Pi\bare)_e|\ge t)&\le2\exp\!\Big(-\frac{2t^2}{\sum_{r,j}(2|\Pi_{ej}|b/m)^2}\Big)\\
&=2\exp\!\Big(-\frac{mt^2}{2b^2\sum_j\Pi_{ej}^2}\Big),
\end{align*}
and $\sum_j\Pi_{ej}^2=(\Pi^2)_{ee}=\Pi_{ee}\le\rho$. A union bound over $e$ gives the first claim. Take $t=\Delta/(2L)$ and apply Lemma~\ref{lem:regret}. For the bias case, $\|\hat c_\Pi-c\|_\infty\le\|\Pi\bare\|_\infty+\beta$, so it suffices that $\|\Pi\bare\|_\infty<\Delta/(4L)$ when $\beta\le\Delta/(4L)$; use $t=\Delta/(4L)$.
\end{proof}

\begin{proof}[Proof of Theorem~\ref{thm:learned}]
Let $\hat\Pi$ and $\hat c$ be computed from the training set, independent of the test pair. Lemma~\ref{lem:decomp} with $\bar c'=\hat c$ gives $\|\hat c_{\hat\Pi}-c\|^2=\|\hat\Pi\bare\|^2+\|(I-\hat\Pi)(c-\hat c)\|^2$. Given the training set, $\E\|\hat\Pi\bare\|^2=\tr(\hat\Pi)s^2=ks^2$. Next, $(I-\hat\Pi)(c-\hat c)=(I-\hat\Pi)(c-\bar c)+(I-\hat\Pi)(\bar c-\hat c)$, and $c-\bar c\in\operatorname{range}\Pi$ where $\Pi$ projects on the range of $\Sigma_c$, so $\|(I-\hat\Pi)(c-\bar c)\|\le\|(I-\hat\Pi)\Pi\|\,\|c-\bar c\|=\sin\Theta\,\|c-\bar c\|$. Hence $\E\|(I-\hat\Pi)(c-\hat c)\|^2\le2\sin^2\Theta\,\tr\Sigma_c+2\|\hat c-\bar c\|^2$. The top-$k$ eigenspace of $\Sigma_y$ is $\operatorname{range}\Pi$ with eigengap $\lambda_k$, so the Davis--Kahan bound in the form of \citet[Cor.~1]{yu2015} gives $\sin\Theta\le2\|\hat\Sigma-\Sigma_y\|/\lambda_k$. For sub-Gaussian vectors and $n\ge d$, with probability at least $1-2e^{-d}$ we have $\|\hat\Sigma-\Sigma_y\|\le C_0K^2\|\Sigma_y\|\sqrt{d/n}$ \citep[Thm.~4.7.1 and Exercise~4.7.3]{vershynin2018}, where centering at $\hat c$ instead of $\bar c$ changes $\hat\Sigma$ by a rank-one matrix of norm $\|\hat c-\bar c\|^2$ of lower order, absorbed in $C_0$. With $\|\Sigma_y\|=\lambda_1+s^2$, squaring gives the second term of the bound with $C=8C_0^2$. Finally $\E\|\hat c-\bar c\|^2=\tr\Sigma_y/n$.
\end{proof}

\begin{proof}[Proof of Proposition~\ref{prop:hybrid}]
Write $\hat c^{\rm hyb}-c=\bar r+\Pi(r-\bar r+\bare)-r=\Pi\bare-(I-\Pi)(r-\bar r)$. The same orthogonality and expectation steps as in Theorem~\ref{thm:mse} give the error, using $\E[\bare\mid c,x]=0$. The predictor $h(x)+\bar r$ has error $\E\|r-\bar r\|^2=\tr\Sigma_r$. The difference is $\tr(\Pi D)/m-\tr(\Pi\Sigma_r)$. The last statement is Theorem~\ref{thm:mse} with $\Sigma_r$ in place of $\Sigma_c$, since its optimal projector is no worse than $\Pi=0$ (the predictor) and $\Pi=I$ (the sample mean, whose error is $\tr D/m$).
\end{proof}

\section{Additional experiments (comparison with SPO+)}

The following results use the SPO+ training protocol from Section~\ref{sec:exp} at $m=1$.

\begin{table}[ht]
\centering
\caption{Paired differences in mean regret (percentage points) and $t$-test $p$-values, $n=200$, $100$ experiment replications. A negative number favors the first method. $^\dagger$: not significant after Bonferroni correction.}
\label{tab:tests}
\footnotesize
\begin{tabular}{@{}lccc@{}}
\toprule
Comparison & $\deg=1$ & $\deg=2$ & $\deg=4$ \\
\midrule

SPO+ $-$ Noisy (naive Dijkstra) & $-$5.89 ($<$0.001) & $-$4.48 ($<$0.001) & +3.47 ($<$0.001) \\
PCA, $k{=}5$ $-$ SPO+ & +0.66 ($<$0.001) & +0.49 ($<$0.001) & +4.34 ($<$0.001) \\
Hybrid, GD $\hat k$ $-$ SPO+ & $-$0.25 ($<$0.001) & $-$0.14 ($<$0.001) & $-$1.57 ($<$0.001) \\
Hybrid, GD $\hat k$ $-$ PCA, $k{=}5$ & $-$0.92 ($<$0.001) & $-$0.63 ($<$0.001) & $-$5.91 ($<$0.001) \\
Hybrid, GD $\hat k$ $-$ Noisy (naive Dijkstra) & $-$6.14 ($<$0.001) & $-$4.63 ($<$0.001) & +1.90 ($<$0.001)\\

\bottomrule
\end{tabular}
\end{table}

\begin{table}[ht]
\centering
\caption{Mean regret (\%) with $n=100{,}000$ training observations, $10$ experiment replications, $1{,}000$ test instances. Best per column in bold.}
\label{tab:large}
\small
\begin{tabular}{@{}lccc@{}}
\toprule
Method & $\deg=1$ & $\deg=2$ & $\deg=4$ \\
\midrule
Noisy (naive Dijkstra) & 6.735 $\pm$ 0.131 & 5.880 $\pm$ 0.178 & 4.485 $\pm$ 0.409 \\
SPO+ & 0.029 $\pm$ 0.004 & \textbf{0.630 $\pm$ 0.218} & 3.719 $\pm$ 0.482 \\
PCA, $k{=}5$ & 1.121 $\pm$ 0.080 & 1.856 $\pm$ 0.219 & 9.634 $\pm$ 0.791 \\
PCA, GD $\hat k$ & 1.121 $\pm$ 0.080 & 1.758 $\pm$ 0.193 & 3.852 $\pm$ 0.399 \\
Ledoit-Wolf filter & 1.112 $\pm$ 0.079 & 1.364 $\pm$ 0.130 & \textbf{2.002 $\pm$ 0.161} \\
Hybrid, GD $\hat k$ & \textbf{0.009 $\pm$ 0.002} & 0.991 $\pm$ 0.230 & 5.737 $\pm$ 1.381\\
\bottomrule
\end{tabular}
\end{table}

\begin{table}[ht]
\centering
\caption{Selected rank and variance explained (means over experiment replications, $m=1$, $n=200$).}
\label{tab:khat}
\small
\begin{tabular}{@{}lccc@{}}
\toprule
 & $\deg=1$ & $\deg=2$ & $\deg=4$ \\
\midrule
GD $\hat k$, raw costs & 5.0 & 5.1 & 12.0 \\
GD $\hat k$ on SPO+ residuals (hybrid) & 4.9 & 5.0 & 11.9 \\
\midrule
Var.\ explained, $k{=}2$, noisy (\%) & 32.7 & 41.4 & 39.6 \\
Var.\ explained, $k{=}5$, noisy (\%) & 61.2 & 77.0 & 72.3 \\
Var.\ explained, $k{=}5$, noise-free (\%) & 100.0 & 95.7 & 79.6 \\
Effective rank, noisy & 19.4 & 12.4 & 12.4 \\
Effective rank, noise-free & 4.7 & 5.8 & 8.9\\
\bottomrule
\end{tabular}
\end{table}

\begin{figure}[ht]
\centering
\includegraphics[width=0.78\textwidth]{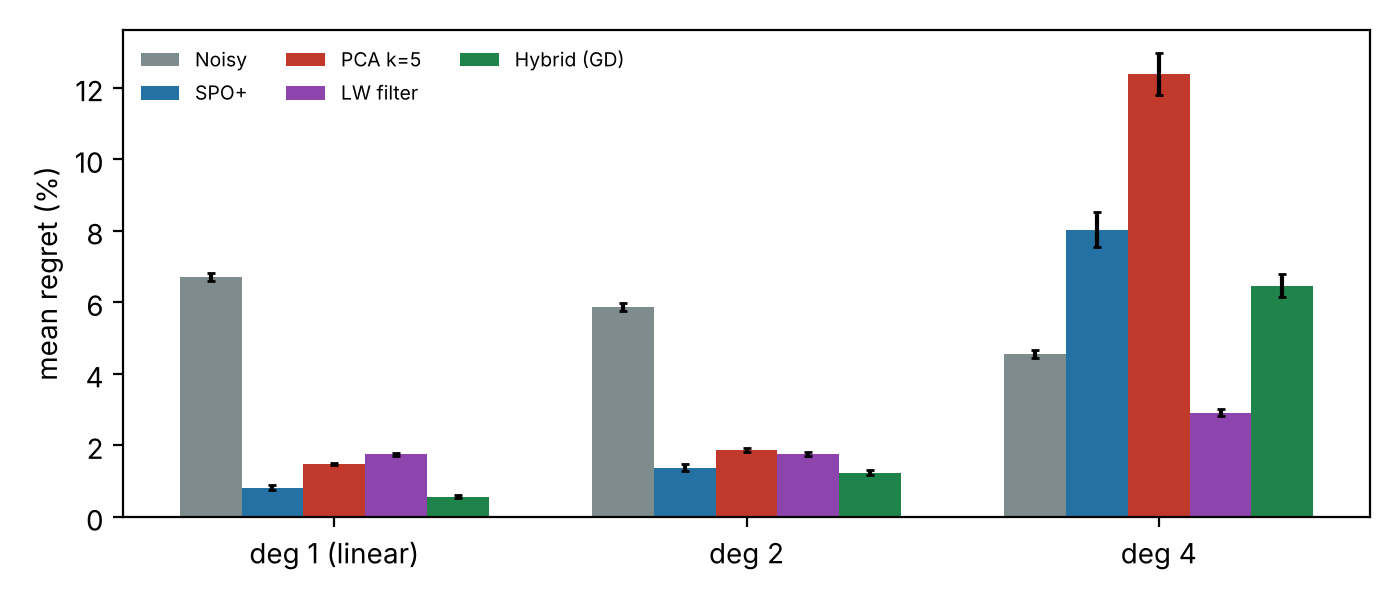}
\caption{Mean regret with $95\%$ confidence intervals ($100$ experiment replications, $m=1$, $n=200$).}
\label{fig:regret}
\end{figure}

\begin{figure}[ht]
\centering
\includegraphics[width=\textwidth]{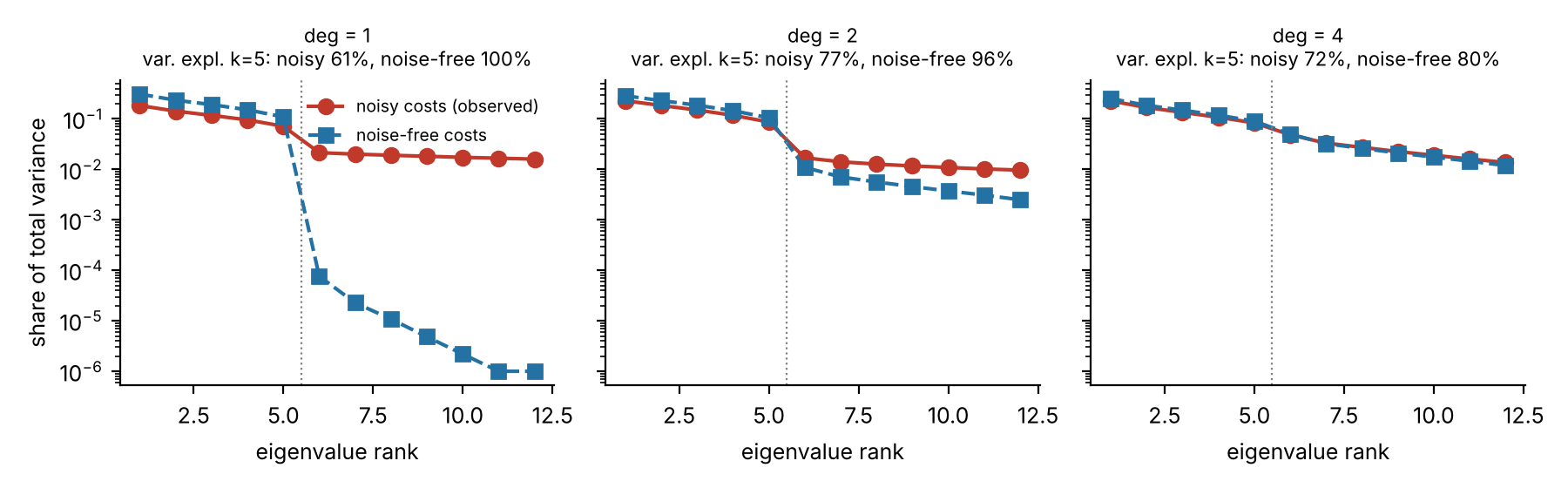}
\caption{Eigenvalue spectra of the observed and noise-free training cost covariance. The dotted line marks $k=5$.}
\label{fig:spectra}
\end{figure}

\subsection{Training Size}
\label{sec:nsweep}

\begin{figure}[ht]
\centering
\includegraphics[width=\textwidth]{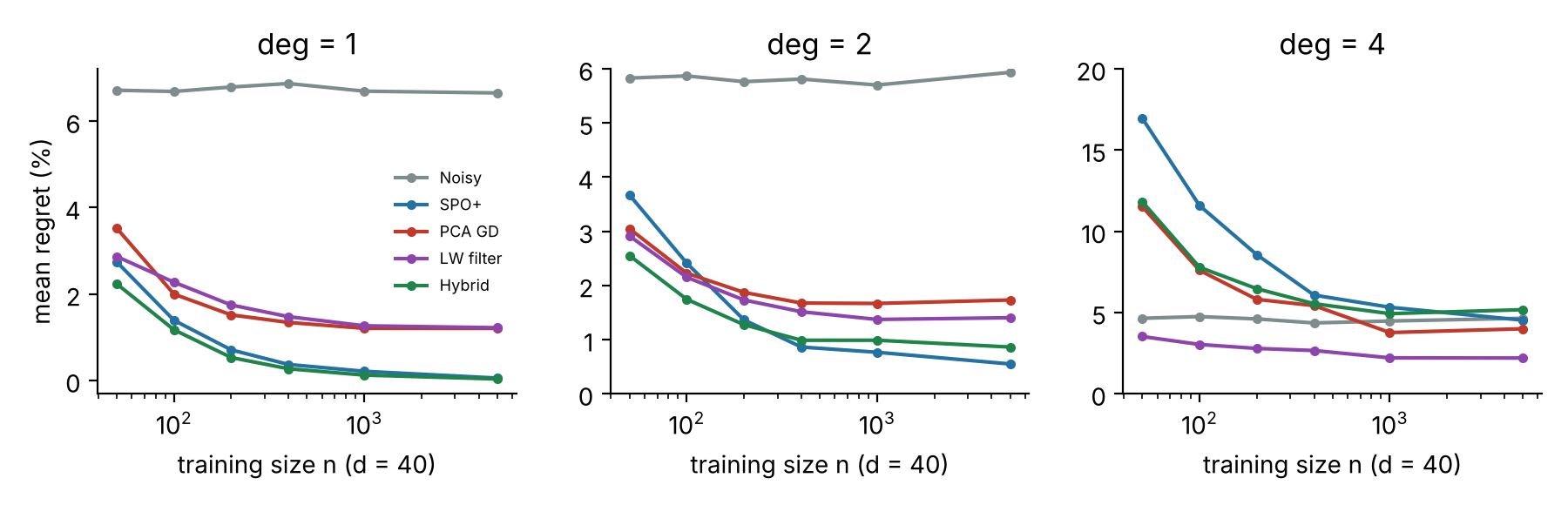}
\caption{Mean regret against training size $n$ ($20$ replications, $d=40$).}
\label{fig:nsweep}
\end{figure}

Table~\ref{tab:nsweep} and Figure~\ref{fig:nsweep} show regret as the training
sample grows from $n=50$ to $n=5000$.

\begin{table}[ht]
\centering
\caption{Mean regret (\%) by training size $n$ ($d=40$, $20$ replications). Best of the four methods shown in bold.}
\label{tab:nsweep}
\small
\begin{tabular}{@{}rrcccc@{}}
\toprule
$\deg$ & $n$ & SPO+ & PCA GD & LW & Hybrid \\
\midrule
1 & 50 & 2.75 & 3.53 & 2.87 & \textbf{2.24} \\
 & 200 & 0.71 & 1.52 & 1.75 & \textbf{0.53} \\
 & 1000 & 0.21 & 1.21 & 1.27 & \textbf{0.13} \\
 & 5000 & 0.06 & 1.21 & 1.22 & \textbf{0.04} \\
\midrule
2 & 50 & 3.67 & 3.05 & 2.91 & \textbf{2.55} \\
 & 200 & 1.36 & 1.87 & 1.73 & \textbf{1.27} \\
 & 1000 & \textbf{0.77} & 1.67 & 1.37 & 0.99 \\
 & 5000 & \textbf{0.55} & 1.73 & 1.40 & 0.86 \\
\midrule
4 & 50 & 16.96 & 11.51 & \textbf{3.52} & 11.82 \\
 & 200 & 8.56 & 5.81 & \textbf{2.79} & 6.46 \\
 & 1000 & 5.32 & 3.77 & \textbf{2.21} & 4.93 \\
 & 5000 & 4.53 & 4.00 & \textbf{2.20} & 5.17\\
\bottomrule
\end{tabular}
\end{table}

\emph{Linear case.} The hybrid has the lowest regret at every sample size and beats SPO+ at each (paired $p<0.01$). The absolute gap shrinks with $n$, from $0.51$ points at $n=50$ to $0.02$ at $n=5000$. The relative gap stays between about $15\%$ and $45\%$. SPO+ improves steadily with $n$ ($2.75\%$ to $0.06\%$). PCA alone stalls near $1.2\%$. PCA alone matches SPO+ only at the smallest sample, $n=50$ ($n/d=1.25$), where the difference is not significant
($p=0.35$).

\emph{Moderate nonlinearity.} The hybrid beats SPO+ for small
samples ($1.12$ points at $n=50$, $0.67$ at $n=100$). The two are
indistinguishable at $n=200$ and $n=400$. For $n\ge1000$ SPO+ is better
($0.55\%$ against $0.86\%$ at $n=5000$). The crossover is at roughly
$n\approx300$, or $n/d\approx 7$.

\emph{Strong nonlinearity.} Every method that uses a fixed small projection loses to naive Dijkstra for most sample sizes. The hybrid is better than SPO+ at small $n$, and about equal from $n=400$. The Ledoit-Wolf filter is the only method that is clearly better than naive Dijkstra at all sizes.

\subsection{Larger and Non-Grid Networks}
\label{sec:topo}

Table~\ref{tab:topo} gives results on a $10\times10$ grid, a $20\times20$ grid, and a layered non-grid network. Each cell has only $6$ replications, so the $p$-values quoted here are unadjusted and the power is low.

\begin{table}[ht]
\centering
\caption{Mean regret (\%) on larger networks ($6$ replications). Best of the five methods shown in bold.}
\label{tab:topo}
\small
\setlength{\tabcolsep}{5pt}
\begin{tabular}{@{}lrrccccc@{}}
\toprule
Network & $n$ & $\deg$ & Noisy & SPO+ & PCA GD & LW & Hybrid \\
\midrule
$10{\times}10$ grid ($d{=}180$) & 200 & 1 & 7.63 & 1.35 & 0.74 & 1.66 & \textbf{0.47} \\
 &  & 2 & 6.88 & 1.97 & 1.68 & 1.92 & \textbf{1.49} \\
 &  & 4 & 5.52 & 11.67 & 4.18 & \textbf{3.14} & 5.03 \\
 & 1000 & 1 & 7.83 & 0.28 & 0.38 & 0.83 & \textbf{0.10} \\
 &  & 2 & 6.75 & \textbf{0.89} & 0.95 & 0.93 & 0.95 \\
 &  & 4 & 5.56 & 6.58 & 2.87 & \textbf{2.07} & 4.22 \\
\midrule
$20{\times}20$ grid ($d{=}760$) & 200 & 1 & 8.81 & 1.62 & 0.55 & 2.16 & \textbf{0.45} \\
 &  & 2 & 7.55 & 2.45 & 1.24 & 2.35 & \textbf{1.17} \\
 &  & 4 & 5.91 & 15.79 & \textbf{4.43} & 4.55 & 5.93 \\
 & 1000 & 1 & 8.69 & 0.38 & 0.19 & 1.38 & \textbf{0.11} \\
 &  & 2 & 7.59 & 1.12 & 0.49 & 1.45 & \textbf{0.47} \\
 &  & 4 & 5.85 & 7.48 & \textbf{2.60} & 3.57 & 4.40 \\
\midrule
Layered DAG ($d{=}192$) & 200 & 1 & 11.27 & 5.19 & \textbf{1.09} & 2.48 & 1.29 \\
 &  & 2 & 9.41 & 5.91 & 2.97 & 3.53 & \textbf{2.75} \\
 &  & 4 & \textbf{6.81} & 46.13 & 27.15 & 15.61 & 30.95 \\
 & 1000 & 1 & 11.28 & 3.04 & \textbf{0.56} & 1.20 & 0.68 \\
 &  & 2 & 9.41 & 2.73 & 2.11 & \textbf{1.84} & 2.51 \\
 &  & 4 & \textbf{7.05} & 14.02 & 15.10 & 8.99 & 24.41\\
\bottomrule
\end{tabular}
\end{table}

\emph{PCA alone beats SPO+ in the linear case on larger networks.} On
the $20\times20$ grid ($d=760$), PCA reaches $0.55\%$ at $n=200$ against $1.62\%$ for SPO+ ($p=0.002$), and $0.19\%$ against $0.38\%$ at $n=1000$ ($p=0.002$). On the layered network ($d=192$), the gap is larger: $1.09\%$ against $5.19\%$ at $n=200$ and $0.56\%$ against $3.04\%$ at $n=1000$. On the $10\times10$ grid, PCA wins at $n=200$ ($0.74\%$ against $1.35\%$) and ties at $n=1000$ ($0.38\%$ against $0.28\%$, $p=0.21$). On the $5\times5$ grid, PCA alone did not beat SPO+ except at $n=50$, where the difference was not significant. In all the winning cells, the sample is small relative to the number of edges ($n/d$ of about $5$ or less). SPO+ must fit $d\times p$ weights, while PCA exploits the low rank of the cost matrix. We have not tested whether that explains the pattern.

\emph{The hybrid improves on both in the linear case on grids.} It beats
SPO+ on every network and sample size in the linear case ($p<0.05$ in all six cells, and $p\le0.002$ in five). It beats PCA alone on both grids, at both sample sizes, in all $6$ replications. On the layered network, PCA alone is at least as good as the hybrid ($1.09\%$ against $1.29\%$ at $n=200$, $p=0.054$).

\emph{Nonlinear costs.} At $\deg=2$ the hybrid is best or tied on the
$20\times20$ grid ($1.17\%$ and $0.47\%$), and SPO+ is best on the
$10\times10$ grid at $n=1000$ ($0.89\%$). At $\deg=4$ on grids, the GD-rank PCA
denoiser and the LW filter beat naive Dijkstra, SPO+ does not, and the GD-rank hybrid
beats naive Dijkstra in three of the four grid cells and ties it in the fourth ($n=200$ on the $20\times20$ grid). On
the layered network at $\deg=4$ every method is worse than naive Dijkstra, at both sample
sizes. We have not isolated the cause. One untested candidate is that paths
on this network differ in length. On a grid, a common shift in the cost level
cannot change the best path, because all paths have the same number of
edges. On the layered network it can.

\subsection{Computation}

\begin{table}[ht]
\centering
\caption{Time (seconds, median of $5$) for a full eigendecomposition of the $d\times d$ covariance against a truncated SVD ($k=5$) of the $200\times d$ data matrix. The last column is the sine of the largest principal angle between the two $5$-dimensional subspaces.}
\label{tab:timing}
\small
\begin{tabular}{@{}rcccc@{}}
\toprule
$d$ & Full eigh & svds ($k{=}5$) & Speedup & Subspace gap \\
\midrule
40 & 0.0002 & 0.0008 & 0.2$\times$ & 2.1e-08 \\
180 & 0.0024 & 0.0012 & 2.0$\times$ & 3.9e-08 \\
760 & 0.0564 & 0.0048 & 11.7$\times$ & 0.0e+00 \\
3120 & 2.6888 & 0.0106 & 252.7$\times$ & 3.3e-08\\
\bottomrule
\end{tabular}
\end{table}

The truncated solver is slower at $d=40$, where a dense solve costs
microseconds. It is faster by $2.0\times$ at $d=180$, $12\times$ at $d=760$, and over $250\times$ at $d=3120$. The two subspaces agree to about $10^{-8}$.
The hybrid needs one extra pass, an SPO+ training run, which dominates the cost
at large $n$.

\end{document}